\documentclass[11pt]{PoS}
\usepackage[T1]{fontenc}
\usepackage[utf8]{inputenc}
\usepackage{amsmath,amssymb,amsfonts}

\title{Studying the gradient flow coupling in the Schr\"odinger functional}
\ShortTitle{Studying the gradient flow coupling in the SF}

\author{\hfill\parbox{3cm}{\it%
DESY 13-150   \\
HU-EP-13/38   \\ 
SFB/CPP-13-57 \\
}}
\author{%
\speaker{P.~Fritzsch}\\
Institut~f\"ur~Physik, Humboldt-Universit\"at~zu~Berlin, \\
Newtonstr.~15, 12489~Berlin, Germany\\
E-mail:~\email{fritzsch@physik.hu-berlin.de}
}
\author{%
A.~Ramos\\
NIC, DESY, \\
Platanenallee~6, 15738~Zeuthen, Germany\\
E-mail:~\email{alberto.ramos@desy.de}%
}
\abstract{%
\vskip1em
We discuss the setup and features of a new definition of the running
coupling in the Schr\"odinger functional scheme based on the gradient 
flow. Its suitability for a precise continuum limit in QCD is demonstrated 
on a set of $\nf=2$ gauge field ensembles in a physical volume of 
$L\sim0.4 {\rm fm}$. 
}

\FullConference{The 31th International Symposium on Lattice Field Theory - Lattice 2013,\\
		July 29 -- August 03, 2013\\
		Mainz, Germany}

\def\gbsq{\overline{g}^2}
\def\gbar{\overline{g}}

\def\vecp{{\bf p}}
\def\nf{N_{\rm f}}
\def\Or{{\rm O}}
\def\fm{{\rm fm}}

\def\cMagenta{\color[cmyk]{0,1,0,0}} 
\def\cRedDark{\color[rgb]{0.75,0,0}}
\begin{document}

\section{Introduction}

Some years ago the Yang-Mills gradient flow was introduced to the Lattice Field
Theory community by M.L\"uscher~\cite{Luscher:2010iy} as an additional tool for 
studying the dynamics of gauge theories non-perturbatively.
Since then the number of applications of the YM gradient flow to
probe non-perturbative aspects of non-Abelian gauge theories using old and new
ideas steadily increases --- as can be seen in the various contributions to
this conference for instance. 
So far most of them are influenced by~\cite{Luscher:2010iy} and the all-order
proof in perturbation theory given in~\cite{Luscher:2011bx} which tells us that
the gauge field that is generated by the flow equations 
\begin{subequations}
\begin{align}\label{eq:DEQ}
    \dfrac{ {\rm d}B_\mu(x,t)}{{\rm d}t} &= D_\nu G_{\nu\mu}(x,t) \;, &  t&> 0 \;,\\ 
                         G_{\mu\nu}(x,t) &= \partial_\mu B_\nu - \partial_\nu B_\mu + [B_\mu,B_\nu] \,,  &  D_\nu &= \partial_\nu + [B_{\nu},\star] \;, \\[0.2em] 
                       B_\mu(x,t)|_{t=0} &= A_\mu(x)  \,,    \label{eq:IC}
\end{align}
\end{subequations}
does not require renormalization. Here $A_{\mu}(x)$ is the fundamental gauge
field of the underlying theory and the flow field $B_{\mu}(x,t)$ is the
solution to \eqref{eq:DEQ} subject to the initial condition~\eqref{eq:IC}.  
On the lattice the gradient flow is also referred to as Wilson flow if the
Wilson plaquette gauge action has been used to define the flow equations in
terms of parallel transporters~\cite{Luscher:2009eq,Luscher:2010iy}. In that
case it has rigorously been shown that the action is a monotonically decreasing
function of $t$ and the flow thus represents a smoothing operation of the
initial gauge field.

\section{Definition of a new coupling}

Here we are especially interested in one of the applications that were already
mentioned in~\cite{Luscher:2010iy}, namely to use the energy density 
\begin{align}
        \left\langle E(t) \right\rangle &\equiv \frac{1}{4} \left\langle {G}_{\mu\nu}(t){G}_{\mu\nu}(t) \right\rangle
     =  \dfrac{3(N^2-1)}{2(8\pi t)^{2}}  \times \gbsq_{\rm MS}(\mu) \Big\{  1 + c_1 \gbsq_{\rm MS} + \Or\big(\gbar_{\rm MS}^4\big)  \Big\}
\end{align}
at positive flow time to define a running coupling. This perturbative expansion
of the energy density in terms of a renormalized coupling is given for 4
Euclidean space-time dimensions in infinite volume for gauge group $SU(N)$ at
scale $\mu=1/\sqrt{8t}$, where $\sqrt{8t}$ is the mean-square radius over which
the gauge field is effectively smoothed.  This obviously can serve as a
definition of a non-perturbatively renormalized coupling after switching from
dimensional regularisation to the lattice as a regulator, leading to the
gradient flow coupling
\begin{align}
    \left\langle t^2 E(t) \right\rangle   &\equiv  \mathcal{N}  \times \gbsq_{\rm GF}(\mu) \;, 
     & \mu&=1/\sqrt{8t} = 1\big/ cL \;.
\end{align}
The energy density as a gauge invariant quantity scales $\propto t^{-2}$
and is a renormalized quantity at positive flow time. The normalization
constant $\mathcal{N}$ has to be chosen such that $\gbsq_{\rm GF} = g_0^2 +
\Or(g_0^4)$ holds and can be computed by expanding the energy density in the
bare gauge coupling as defined through the field strength tensor
$G_{\mu\nu}(t)$.  In a finite volume one has the additional length scale $L$,
the physical size of a hypercube with volume $V=L^4$, which in a finite size
scaling procedure usually sets the renormalization scale in some well-defined
way. It is necessary to fix the factor $c=\sqrt{8t}/L$ that defines the 
effective smoothing range in terms of the physical extent.

In a finite volume, boundary conditions of the field variables become important
and the energy density has in fact already been used to define a running coupling
with periodic boundary conditions~\cite{Fodor:2012td}.  Unfortunately, this
leads to a definition of the coupling that is non-analytic in $g_0^2$ and thus
has a non-universal 2--loop coefficient in the QCD $\beta$-function. It
considerably complicates the perturbative computations beyond tree-level that 
are needed to safely relate the $\Lambda$-parameter of this scheme to 
$\Lambda_{\overline{\rm MS}}$.

As it is known for a long time, this behaviour can be avoided from the start by either
using twisted boundary conditions or the Schr\"odinger functional as
finite-volume renormalization scheme~\cite{Luscher:1992an} where Dirichlet
boundary conditions are imposed at the time boundaries.  One of the authors was
also working on a computation of the gradient flow coupling using twisted
boundary conditions and presented its results for the full step-scaling
function in $SU(2)$ pure gauge theory~\cite{Alberto:Lat13} that shows a
promising accuracy towards a new computation of the running coupling and the
Lambda parameter in QCD.

\section{The gradient flow coupling in the SF}

The Schr\"odinger functional (SF) is the Euclidean propagation kernel of some
field configuration at Euclidean time $x_0=0$ to $x_0=T$ where $T$ is the
extent of the finite-volume world in time direction.  In the spatial direction,
gauge fields have periodic boundary conditions with period $L$ while Dirichlet
boundary conditions are imposed to the spatial components of the gauge fields
at $x_0=0,T$. Accordingly time translation invariance is lost and all physical
observables explicitly depend on the Euclidean time. In our setup we are only
considering vanishing boundary fields. This means that for a spatial Fourier
transformed flow field, $\tilde B_{\mu}(\mathbf{p},x_0,t)$, one would impose
\begin{align}
  \forall\mathbf{p}: &&             \tilde B_k(\mathbf{p}, x_0, t)\big|_{x_0=0,T}            &=0   \;,
\intertext{while the boundary condition of the time component emerges through the gauge fixing procedure,}
  \mathbf{p} \ne 0 : &&  \partial_0 \tilde B_0(\mathbf{p}, x_0, t)\big|_{x_0=0,T}            &=0   \;, \\
  \mathbf{p}   = 0 : &&             \tilde B_0(\mathbf{0}, x_0, t)\big|_{x_0=0\hphantom{,T}} &=0   \;,
                      &  \partial_0 \tilde B_0(\mathbf{0}, x_0, t)\big|_{x_0=T}              &=0   \;. \label{eq:cont-B0-p=0}
\end{align}
This is best seen by starting from the lattice formulation and taking the
continuum limit. For additional details we have to refer the interested reader
to our paper~\cite{Fritzsch:2013je} and references therein. 

As mentioned earlier the normalization factor $\mathcal{N}$ is obtained by
expanding the flow field and thus the energy density in terms of the bare
coupling. At $t>0$ this reads
\begin{align}
        B_{\mu} &= \sum_{n=1}^{\infty} B_{\mu,n} g_0^{n} \;, &
        \langle  E(t,x_0) \rangle &=  \sum_{n=0}^{\infty} \mathcal E_n(t,x_0)  \;,~\text{with}\quad\mathcal E_n=\Or\big(g_0^{2+n}\big) \;,\\
        && {\cMagenta\text{LO:}}\quad
  \mathcal E_0(t,x_0) &= \frac{g_0^2}{2}\langle
  \partial_{\mu}B_{\nu,1}^a\partial_{\mu}B_{\nu,1}^a - 
  \partial_{\mu}B_{\nu,1}^a\partial_{\nu}B_{\mu,1}^a
  \rangle \;,   \label{eq:E0}
\end{align}
and only the leading order (LO) will contribute to $\mathcal{N}$.  Inserting
the expansion of $B_{\mu}$ into the flow equation reads to leading order
\begin{align}
    \dfrac{{\rm d}}{{\rm d}t} \tilde B_{\mu,1}(\mathbf p, x_0, t) &=  (-\mathbf p^2 + \partial_0^2) 
    \tilde B_{\mu,1}(\mathbf p, x_0, t) \;,& 
    \tilde B_{\mu,1}(\mathbf p, x_0, t)\big|_{t=0} &= \tilde A_{\mu}(\mathbf p, x_0) \;,
\end{align}
which evidently is a heat equation. Solutions to these kind of equations 
have been known for a long time and can be written in terms of heat kernels.
Those that enter our observable can be written as
\begin{align}
    \tilde B_{\mu,1}(\vecp,x_0,t) &= {\rm e}^{-\vecp^2 t} \int^T_0\!\!{\rm d}x_0^\prime\, K^{D,N}(x_0,x_0^\prime,t) \, \tilde A_{\mu}(\vecp,x_0^\prime,t) 
     \;,\qquad (\mathbf{p}\ne 0)
\end{align}
where $K^{D}$, $K^{N}$ are heat kernels consistent with {\it Dirichlet} 
$(\mu=1,2,3)$ and {\it Neumann} $(\mu=0)$ boundary conditions, respectively, 
at $x_0=0,T$. Inserting these into eq.~\eqref{eq:E0}
leads to partial derivatives of the heat kernels convoluted with the standard
gluon propagator w.r.t. the fundamental gauge fields $A_{\mu}$. Since the
latter is known analytically everything can be written down in closed form and
the normalization factor be read-off. In general the continuum normalization
factor $\mathcal{N}$ depends on the parameter $c$, the time-slice used in
the evaluation of $E(x_0,t)$ and the geometry that is applied, i.e., the ratio 
$T/L$. It is most advantageous to work at $x_0=T/2$ with $T=L$. Useful choices 
for $c$ will be discussed below.
To ease notation and the discussion of the overall computation we worked in the
continuum but the same procedure is carried over to the lattice formulation.
There the corresponding norm $\hat{\mathcal{N}}$ will explicitly depend on the
lattice size $L/a$ and implicitly on further details as the lattice action 
and the actual discretisation of the energy density that has been used.  
In~\cite{Fritzsch:2013je} we have done this computation for the Wilson plaquette 
gauge action and the clover definition of the energy density. Our results have
also been checked by using some dedicated small coupling simulations that
show the correct asymptotic behaviour.

\section{Non-perturbative results}

\begin{figure}[t]
        \centering
        \includegraphics[width=0.75\textwidth]{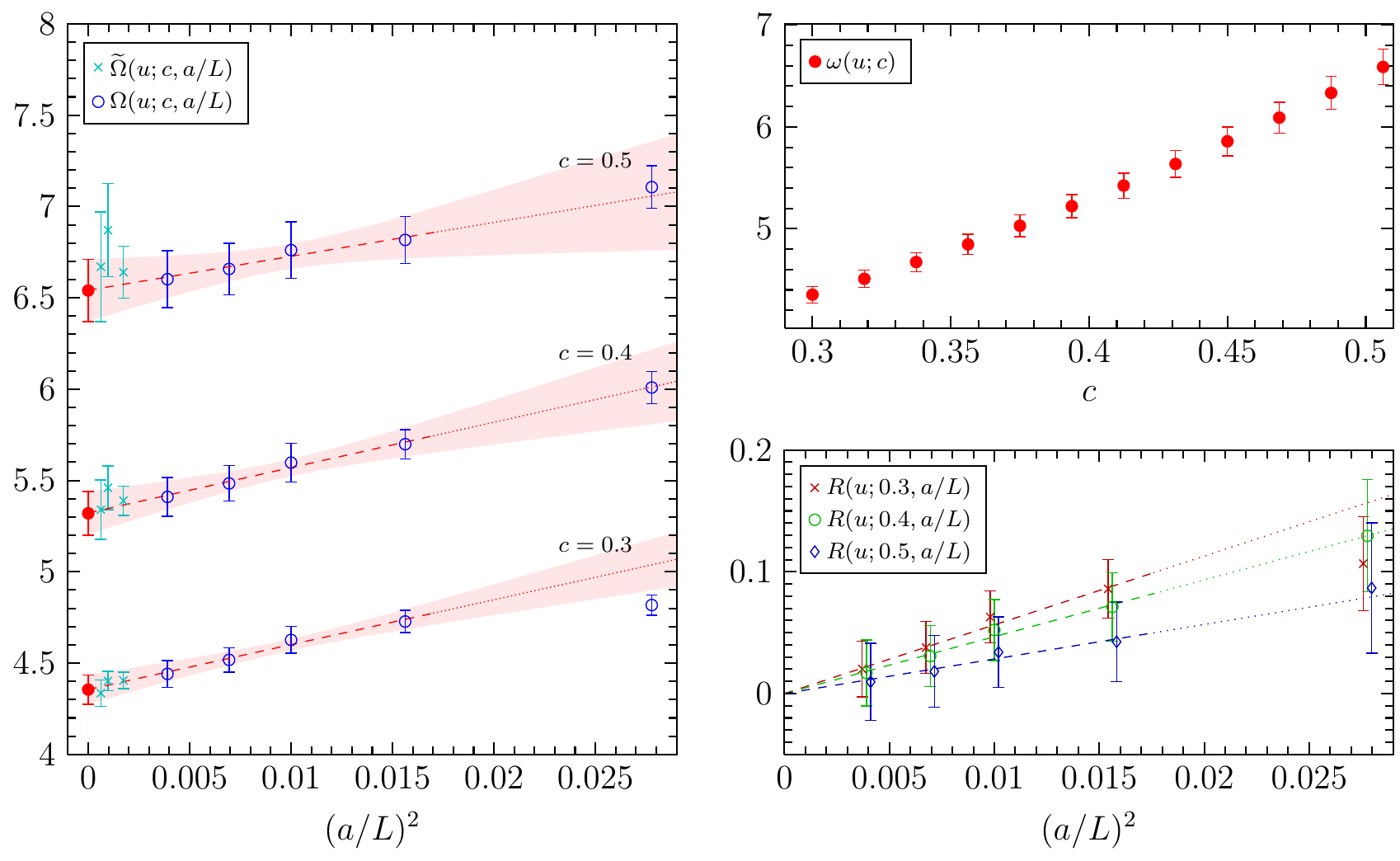}
        \vskip-.75em
        \caption{Exemplary results of $\Omega(u;c,a/L)$ with continuum limits $\omega(u;c)$ and its dependence on $c$.}
        \vskip.25em
        \label{fig:CL}
\end{figure}

To immediately learn more about the general non-perturbative behaviour of the
gradient flow coupling in the SF as defined earlier we decided to compute the
corresponding observable $E(x_0,t)$ for different values of the smearing ratio
$c$ on existing $\nf=2$ gauge field configurations for lattice sizes
$L/a=6,8,10,12,16$ with $T=L$. These have been set up along a trajectory in the
space of bare couplings at which the traditional SF coupling and PCAC quark
mass are fixed to
\begin{align} \label{eq:LCP}
  {\rm \cRedDark LCP:}\qquad
  \overline{g}^{2}_{\rm SF}(L_1) &\equiv u = 4.484  \qquad\text{and}\qquad m(L_1) = 0 \;,
\end{align}
corresponding to a physical box size of $L_1\sim 0.4\,\fm$. Our observable of
interest thus reads
\begin{align}
     \Omega(u;c,a/L) &= 
           \left[ 
           \hat{\mathcal{N}}^{-1}(c,T/L,x_0/T,a/L)\cdot t^2\big\langle
           E(T/2,t)\big\rangle  
           \right]_{\raisebox{0em}{\scriptsize$t=c^2L^2/8$}}^{\raisebox{0em}{\scriptsize{\cRedDark \text{LCP}}}} 
\end{align}
and we show some selected data in figure~\ref{fig:CL}. There the by far
dominating part of the overall error budget ($\gtrsim 85\%$) comes from
propagating the uncertainty of setting up a line of constant physics (LCP)
according to~\eqref{eq:LCP}. The continuum limit is taken without the coarsest
lattice $L/a=6$.  We observe the following: (a) below $c=0.3$ we see deviations
from the naive leading scaling expectation $(a^{-2})$ that gets stronger and
stronger with decreasing smearing range, (b) in the range $0.3\le c \le 0.5$
relative cutoff effects stay below $10\%$ and decrease with increasing $c$, (c)
the uncertainty increases with increasing smoothing range. 

\pagebreak

\vskip1em
\noindent{\bf The relative variance}
\vskip0.5em
In order to judge how accurately the continuum limit can be reached with the SF
gradient flow coupling compared to the traditional SF coupling, we also
computed the relative variance that for any observable $\mathcal{O}$ is given by
\begin{align}
    \mathcal{V_O} &= \dfrac{\langle \mathcal O^2 \rangle - \langle \mathcal O \rangle^2}{\langle \mathcal O \rangle^2} \;.
\end{align}
Since numerical prefactors cancel one has $\mathcal{V}_{\gbsq_{\rm
GF}}=\mathcal{V}_E$.  Note that $\mathcal{V_O}$ is a genuine observable that
tells us about the statistical accuracy that can be achieved for an observables
$\mathcal{O}$ when the continuum limit is approached in contrast to the
integrated autocorrelation time $\tau_{\mathcal{{\rm int},\mathcal{O}}}$ that
tells us how the underlying algorithm performs for this observable.

\begin{figure}[t]
        \centering
        \includegraphics[height=0.2775\textwidth]{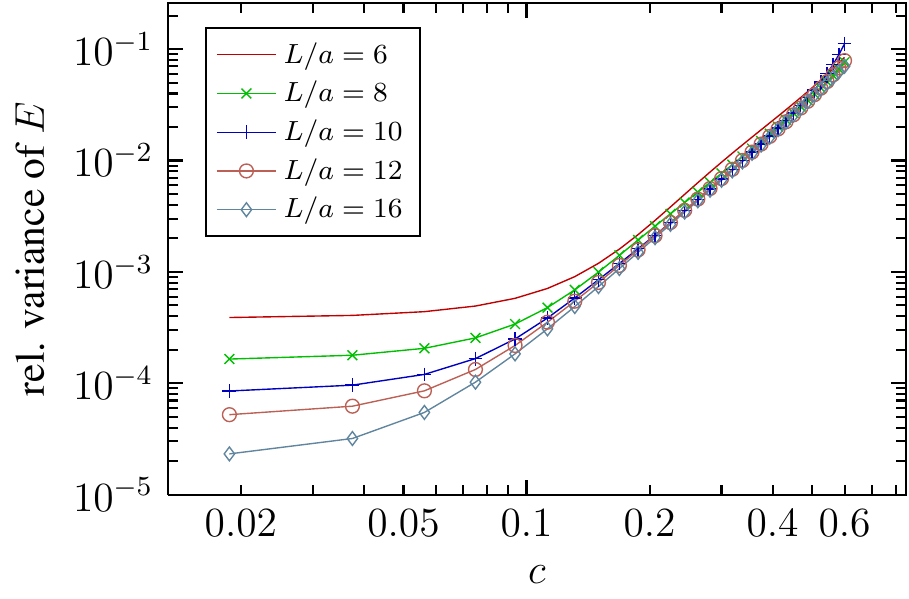}
        \raisebox{0.38em}{\includegraphics[height=0.300\textwidth]{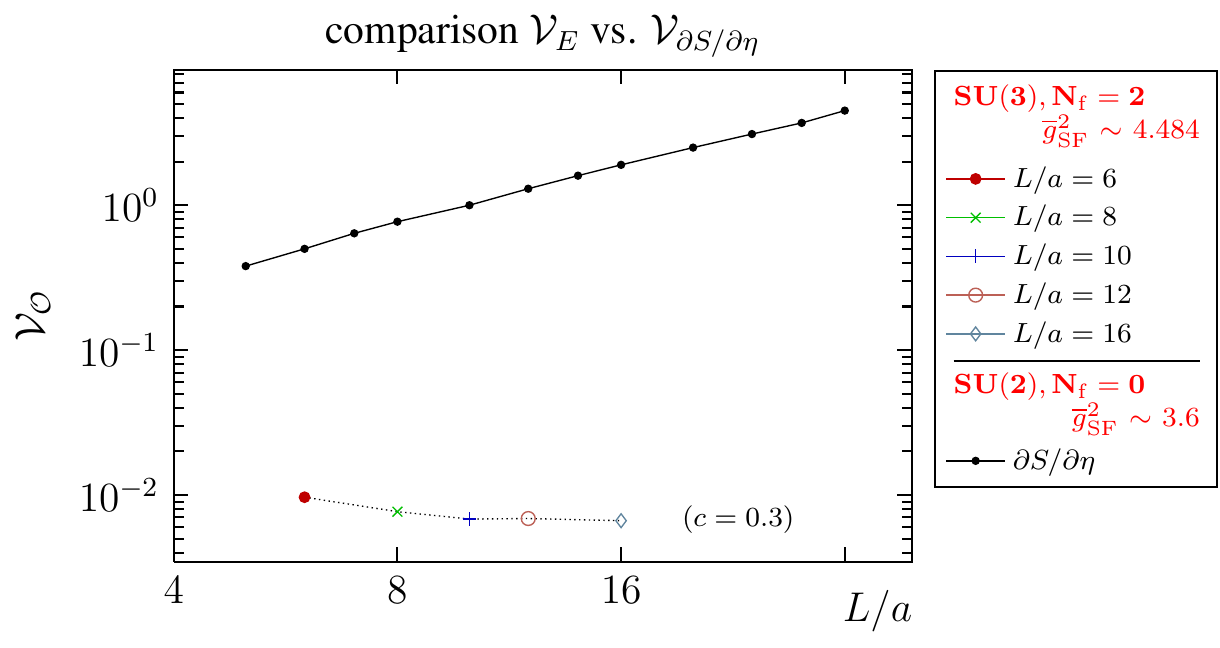}}
        \vskip-.75em
        \caption{Relative variance of the SF gradient flow coupling compared to that of the traditional SF coupling.}
        \label{fig:relvar}
\end{figure}
The results for our five lattices are shown in the left panel of
figure~\ref{fig:relvar}.  As we have noted earlier one wants to avoid ---or is
unable--- to take the continuum limit for $c\ll 0.3$ due to large cutoff
effects and it seems that for say $c>0.2$ the variances $\mathcal{V}_E(a/L)$
fall onto a universal curve.  A similar study has been done for the SF
coupling in the case of pure $SU(2)$ gauge theory~\cite{deDivitiis:1994yz},
showing that the variance of the SF coupling diverges towards the continuum
limit. In the right panel of figure~\ref{fig:relvar} we plot their results from
table~1 together with our result for the SF gradient flow coupling at $c=0.3$.
Note that both are obtained along a line of constant physics defined through a
large but slightly different SF coupling and that the overall behaviour of the
SF coupling is the same in QCD with dynamical flavours as studied in the case
of the gradient flow coupling.
\vskip1em
\noindent{\bf The integrated autocorrelation time of $\gbsq_{\rm GF}$}
\vskip0.5em
Our statistical errors always include an estimate of the integrated
autocorrelation time using the $\Gamma$-method~\cite{Madras:1988ei} and we
observe a scaling of 
\begin{align}
        \left. \tau_{{\rm int},\gbsq_{\rm GF}} \right|_{c=0.3} &\approx (L/a)^2 \times 0.05\,{\rm MDU}
\end{align}
that agrees with the scaling behaviour that is naively expected for a Hybrid
Monte Carlo simulation ($\propto a^{-2}$). At finite flow time also the
topological charge can be measured much more accurately and one could ask how
the different topological sectors couple to $\gbsq_{\rm GF}$, especially in
physically larger lattices, i.e., in the low energy regime of QCD. This question
is directly related to critical slowing down and was studied in more detail
in~\cite{Felix:Lat13}.
\vskip1em
\noindent{\bf Scaling of the numerical Wilson flow integrator scheme}
\vskip0.5em
\begin{figure}[t]
        \centering
        \includegraphics[width=0.5\textwidth]{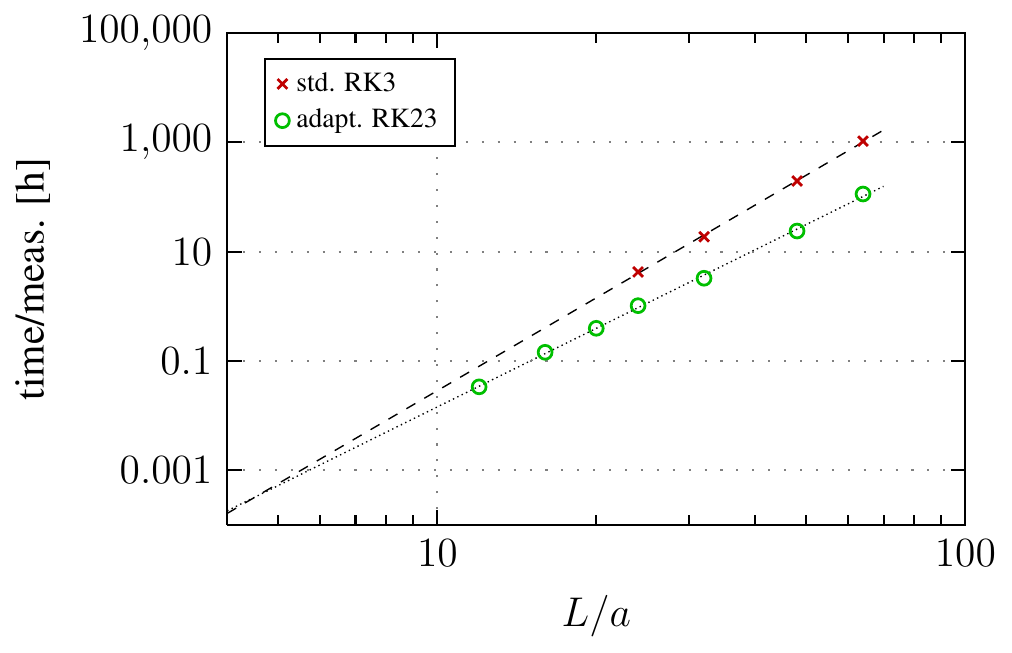}
        \caption{Scaling behaviour of standard Runge-Kutta integrator (RK3)~\cite{Luscher:2010iy} versus
        adaptive step-size integrator (RK23)~\cite{Fritzsch:2013je} for an equivalent setup integrated up 
        to $c_{\rm max}=0.5$}
        \label{fig:intscal}
\end{figure}
In order to integrate the associated flow equations, a first-order differential
equation in the gauge group, the Euler or any Runge-Kutta scheme can be used.  
In~\cite{Fritzsch:2013je} we
extend the originally proposed Runge-Kutta scheme (RK3) with fixed
step-size~\cite{Luscher:2010iy} by nesting a 2nd order scheme to define an adaptive
step size scheme (RK23). Due to the smoothing property of the flow the step size
is safely increased with flow time. For
simulations that we are currently performing with lattices up to $L/a=64$ we
have collected the run time for measurements of Wilson flow observables up to a
fixed flow time and identical setup. The results are plotted in
figure~\ref{fig:intscal} and it can be easily inferred that for the largest
lattice a speed-up factor of $\sim 10$ is seen. Already on a $L/a=32$ lattice,
which may typically be used for the finest resolution in a step-scaling
procedure, a significant speed-up is achieved.

\section{Conclusions \& Outlook}

We have perturbatively computed the continuum and lattice behaviour of the energy 
density at positive gradient flow time in the Schr\"odinger functional with 
vanishing boundary fields~\cite{Fritzsch:2013je}.  This allows us to define a new 
finite-volume renormalization scheme. 
From our studies we see that for wisely chosen flow parameter 
$(0.25\lesssim c \lesssim 0.5)$ a controlled continuum limit can be taken.
Furthermore, we find strong numerical evidence that the new coupling 
can be computed with high accuracy. We also observe that the variance of the
coupling is independent of $L/a$ which will improve continuum determinations 
of observables such as the $\Lambda$-parameter. This new non-perturbative coupling  
may also be very useful in the search for a conformal window in beyond the standard 
model theories.

However, due to the highly improved statistical accuracy there are still many
corners to explore. For instance, in the past perturbatively computed boundary 
improvement terms like $c_{t}$ and $\tilde{c}_t$ were not affecting the results 
or error budget. We are currently investigating such issues with the tree-level 
improved L\"uscher-Weisz gauge action.

\section*{Acknowledgments}
\small

This work is supported in part by the Deutsche Forschungsgemeinschaft under SFB/TR~9.  
We gratefully acknowledge the computer resources provided by the John von Neumann 
Institute for Computing as well as at HLRN and at DESY, Zeuthen.

\bibliography{lattice2}

\end{document}